\documentclass[twocolumn,numberedappendix,twocolappendix,appendixfloats]{openjournal_dhayaa}

\usepackage{amsmath}
\usepackage{fontawesome5}
\usepackage{color}
\usepackage{xcolor}

\usepackage{natbib}
\setcitestyle{aysep={}}

\usepackage[colorlinks=true
  ,urlcolor=blue
  ,anchorcolor=blue
  ,citecolor=blue
  ,filecolor=blue
  ,linkcolor=blue
  ,menucolor=blue
  ,linktocpage=true
  ,pdfproducer=medialab
  ,pdfa=true
]{hyperref}

\newcommand{\Rvir}{R_{\rm vir}}
\newcommand{\Mvir}{M_{\rm vir}}

\newcommand{\cvir}{c_{\rm vir}}

\newcommand{\Om}{\Omega_{\rm m}}
\newcommand{\seight}{\sigma_8}

\newcommand{\eg}{{\sl e.g.}, }

\newcommand{\msol}{\ensuremath{\, {\rm M}_\odot}}    
\newcommand{\msun}{\ensuremath{\, {\rm M}_\odot}} 
\newcommand{\kpc}{\ensuremath{\, {\rm kpc}}}         
\newcommand{\mpc}{\ensuremath{\, {\rm Mpc}}}

\definecolor{orcidlogocol}{HTML}{A6CE39}
\definecolor{purple}{RGB}{128, 0, 128}

\newcommand{\OrcidID}[1]{ \href[urlcolor = red]{https://orcid.org/#1}{\textcolor{lightgray}{\faOrcid}}}
\newcommand{\OrcidIDName}[2]{\href{https://orcid.org/#1}{#2}}

\defcitealias{Anbajagane2022BaryImprint}{A22}
\defcitealias{Shao2023Baryons}{S23}
\defcitealias{Singh2020CosmologyWithMORs}{S20}

\newcommand{\Shao}{Shao2023Baryons}
\newcommand{\Anba}{Anbajagane2022BaryImprint}

\usepackage[T1]{fontenc}
\usepackage{ae,aecompl}

\begin{document}

% Title of the paper, and the short title which is used in the headers.
% Keep the title short and informativxe.
\title{Baryonic Imprints on DM Halos: characterizing the full concentration--mass probability distribution with CAMELS}

\author{\OrcidIDName{0009-0006-1696-2118}{Elisa J. Gao}$^{1, \star}$}
\author{\OrcidIDName{0000-0003-3312-909X}{Dhayaa Anbajagane}$^{1, 2, \star}$}
\email{$^{\star}$elisagjy@uchicago.edu, dhayaa@uchicago.edu}
\affiliation{$^{1}$ Department of Astronomy and Astrophysics, University of Chicago, Chicago, IL 60637, USA}
\affiliation{$^{2}$ Kavli Institute for Cosmological Physics, University of Chicago, Chicago, IL 60637, USA\\}
% \date{\today}

% Abstract of the paper
\begin{abstract}
The halo concentration, which represents the distribution of matter within a halo, is a vital tool for modeling the matter distribution on scales ranging from megaparsecs to sub-kiloparsecs. While significant progress has been made on connecting the impact of astrophysics on the mean concentration as a function of halo mass, the same connection to higher moments of the concentration distribution remains far less explored. We employ the IllustrisTNG runs in the \textsc{Camels} simulation suite to simultaneously quantify the astrophysics and cosmology dependence of the first four moments of the concentration--mass relation; the mean, the scatter, the skewness, and the kurtosis. We use normalizing flows to model the concentration distribution as a function of halo mass, redshift, and six cosmological/astrophysical parameters. We focus on the mass range $\Mvir \in [10^{11}, 10^{14.5}] \msun/h$, and the redshift range $z \in [0, 1]$. All higher-order moments exhibit a mass-dependent sensitivity to both astrophysics and cosmology. The behavior in dark matter-only simulations follows the known effects of active mergers and of tidal stripping in smaller halos. Once galaxy formation is included, the scatter shows non-monotonic behavior at the Milky Way (MW) mass-scale, and the amplitude of this behavior is strongly influenced by the baryon fraction and supernovae (SNe) feedback parameters. The skewness has a large sensitivity to the baryon fraction as well and is mostly insensitive to the other parameters, while the kurtosis follows the behavior set by the skewness. Results for the ASTRID simulation suite show similar behavior as that of IllustrisTNG. The cosmology and astrophysics dependence of these higher-order moments impact analyses that are sensitive to the full distribution of the concentration--mass relation, particularly for halos at/below the MW mass scale.
\end{abstract}

%%%%%%%%%%%%%%%%%%%%%%%%%%%%%%%%%%%%%%%%%%%%%%%%%%

%%%%%%%%%%%%%%%%% BODY OF PAPER %%%%%%%%%%%%%%%%%%

\section{Introduction}

The halo concentration is a single number that represents the shape of the matter distribution within halos \citep{Navarro1997NFWProfile,Bullock2001HaloProfiles,Duffy2008Concentrations,Navarro2010EinastoVsNFW} and is a vital ingredient for modeling many astrophysical and cosmological observables. Using a halo-model prescription \citep{Peacock2000HaloOccupation,Seljak2000HaloModel,Cooray2002HaloModel}, the concentration--mass relation can be used --- in conjunction with a model for the abundance and spatial clustering of halos --- to predict the statistics of the matter density field on quasi-linear and non-linear scales. Such approaches have been verified through extensive validations against predictions from N-body simulations, which numerically solve the full non-linear dynamics of the density field \citep[\eg][]{Seljak2000HaloModel,Mead2015HmCode,Mead2016HmCode,Mead2021HmCode2020,Acuto2021HaloModel}.

However, N-body simulations generally solve for the dynamics of dark matter under gravitational evolution, and do not include gas hydrodynamics nor the plethora of astrophysical effects that impact the evolution of gas, stars, etc. over cosmic time \citep[see][for a review on hydrodynamics simulations]{Vogelsberger2020Hydro}. The latter significantly impacts the halo density profile \citep[\eg][]{Duffy2010BaryonDmProfileDensity, Peirani2017DensityProfile}, with the exact signature depending on the assumed physics of galaxy formation \citep[][]{Kravtsov2012ClusterFormation, Vogelsberger2020Hydro}. This impact has been previously shown by many groups \citep{Duffy2010BaryonDmProfileDensity, Beltz-Mohrmann2021BaryonImpactTNG, Anbajagane2022BaryImprint, Shao2023Baryons} using a wide variety of simulations. The work of \citet[][henceforth, \citetalias{\Shao}]{Shao2023Baryons}, calibrated the concentration--mass relation as a function of cosmology and four hyperparameters associated with the supernovae (SNe) and supermassive black hole (SMBH) physics implemented in the \textsc{Camels} simulations \citep{Villaescusa-Navarro2021CAMELS} under the \textsc{IllustrisTNG}, or TNG, galaxy formation model \citep{Springel2018FirstClustering, Pillepich2018FirstGalaxies, Nelson2018FirstBimodality, Marinacci2018FirstFields, Naiman2018FirstEuropium}.

While the mean relation, $\langle \cvir | \Mvir \rangle$ is a critical ingredient in the aforementioned works, the full distribution $P(\cvir | \Mvir)$ is the more complete prescription. Throughout this work, we use the definitions $\Rvir$, which is the radius within which the average density is $\Delta_{\rm vir} \rho_c$. Here, $\Delta_{\rm vir}$ is given by \citet{Bryan1998vir} and $\rho_c$ is the critical density at a given epoch. The mass $\Mvir$ is then simply the mass within this radius. The concentration probability distribution has been actively propagated into predictions spanning cosmological large-scale structure statistics and observables sensitive to the internal density structure of individual halos. Certain halo-model calculations explicitly average over the scatter in concentration at fixed halo mass \citep{Dolney:2004:ConcentrationScatter, Giocoli:2010:SmallScalePower, Pandey:2025:Godmax} while models for low-mass halo populations in analyses of strong lensing or dark matter annihilation draw concentrations from Gaussian distributions in $\log \cvir$, with a fixed scatter \citep{Fedeli:2007:ConcentrationLensing, Reed:2011:ProfileScatter, Amorisco:2022:ConcentrationLensing, Gilman:2025:SL}. The full distribution, $P(\cvir | \Mvir)$, is therefore already of importance to current modeling infrastructure. This importance will only grow as more precise data is analyzed. While the community has actively studied the mass dependence of the first two moments (the mean and the scatter) of the concentration--mass relation \citep[\eg][]{Wechsler2002Concentrations, Diemer2015Concentration, Anbajagane2022BaryImprint}, the mass dependence of the higher order moments have not been as well characterized. The astrophysical and cosmology dependence of the scatter and the higher moments has also not been well characterized.

This is precisely the measurement we seek to make. The challenge, however, is robustly estimating the function $P(\cvir | \Mvir, z, \vec{\theta})$, where $\vec{\theta}$ is some chosen set of cosmological and astrophysical parameters. In the limit of infinite data, one can simply compute a histogram of $\cvir$ in bins of $\Mvir$, $z$, and $\vec{\theta}$. Given we only have far sparser datasets at our disposal, we can utilize the fact that the distribution is a smooth function of its arguments and employ more optimal techniques than simple binning. This work makes extensive use of normalizing flows, which are neural-network aided methods for modeling entire distributions as a function of some input parameters. We use a combination of 1000 \textsc{Camels} simulations \citep{Villaescusa-Navarro2021CAMELS} to train the distribution $P(\cvir | \Mvir, z, \vec{\theta})$. Here the parameter vector consists of $\vec{\theta} = \{\Omega_{\rm m}, \sigma_8, A_{\rm SN1}, A_{\rm AGN1}, A_{\rm SN2}, A_{\rm AGN2}\}$, i.e. two cosmological parameters, two SNe feedback parameters, and two active galactic nuclei (AGN) feedback parameters. We detail their interpretations further below. Using this distribution, we can compute the first four moments, $\int \cvir ^ n P(\cvir | \ldots )$, and quantify their dependence on astrophysics and cosmology. While we focus primarily on the TNG model, we also comment on comparisons with the ASTRID galaxy-formation model \citep{Bird:2022:ASTRID}, as implemented in the \textsc{Camels}-ASTRID simulation suite \citep{Ni2023Camels28}.

Our work is organized as follows: in Section \ref{sec:DataMethods} we describe the \textsc{Camels} simulation data, including the halo concentration measurements we study, and the normalizing flow methods used to analyze them. We then extract, in Section \ref{sec:DMO}, the cosmology dependence of the four moments of the concentration--mass relation in the simpler dark matter-only (DMO) case. We then present the astrophysics and cosmology dependence in the full hydrodynamics simulation case in Section \ref{sec:hydro}. We conclude in Section \ref{sec:conclusion}. Appendix \ref{appx:linregress} validates our methodology for predicting the full distribution, while Appendix \ref{appx:full_pdf} presents the full probability distribution and Appendix \ref{appx:astrid} briefly discusses the generalizations of this work to other galaxy formation models.

\section{Data and Methods}\label{sec:DataMethods}

\subsection{The \textsc{Camels} simulations}\label{sec:Camels}

The Cosmology and Astrophysics with MachinE Learning Simulations \citep[\textsc{Camels,}][]{Villaescusa-Navarro2021CAMELS}, is a simulation suite of $V = (25 \mpc/h)^3$ boxes, each with varying cosmological and astrophysical parameters. The cosmological parameters include the fractional matter energy density in the present epoch ($\Om$), and the root-mean-square of deviations in the linear density field smoothed on a scale of $8 \mpc/h$ ($\seight$). The astrophysical parameters are $A_{\rm SN1}, A_{\rm SN2}$, which control the supernovae feedback, and $A_{\rm AGN1}, A_{\rm AGN2}$ which control the active galactic nuclei (AGN) feedback. The fiducial values of these parameters are $\Omega_{\rm m} = 0.3$, $\sigma_8 = 0.8$, and $A_{\rm SN1} =  A_{\rm SN2} = A_{\rm AGN1} =  A_{\rm AGN2} = 1$.

We make use of the \textsc{Rockstar} halo catalogs \citep{Behroozi2013Rockstar} from all these simulations. The halo concentration in these catalogs is computed by fitting the total matter density profiles of halos with an NFW form \citep{Navarro1997NFWProfile},
\begin{equation} \label{eqn:NFW_profile}
    \rho(r) = \frac{\rho_s}{\frac{r}{r_s}\big(1 + \frac{r}{r_s}\big)^2}\,,
\end{equation}
where $\rho_s$ and $r_s$ are the characteristic density and the scale radius, respectively. The concentration is then derived as,
\begin{equation}\label{eqn:c200c_def}
    \cvir = \Rvir/r_s,
\end{equation}
Note that the concentration and virial mass here are computed using \textit{all} available particles rather than just the DM alone. Thus, we study the NFW profiles of the total matter distribution and not just the DM distribution alone. Given the DM constitutes 85\% of the matter content, the total matter distribution --- on length scales of interest to us, namely excluding the halo core where the stellar mass distribution is most prominent --- would largely be determined by just the DM distribution.\footnote{Note that \textsc{Rockstar} obtains $\cvir$ by fitting density profiles down to the force softening scale of the simulation, which is $\epsilon \approx2\kpc$ (comoving) for the simulations discussed here. In practice, the smallest informative length scale of the profile is 2-3 times the force softening as densities on scales close to $\epsilon$ are downweighted in the \textsc{Rockstar} fitting procedure.}
The total matter distribution is observationally more relevant than the DM-only distribution, and is conveniently the definition used in these \textsc{Rockstar} catalog.

The \textsc{Camels} simulations were run with $256^3$ dark matter particles of mass $6.5\times 10^7 (\Omega_{\rm m} - \Omega_{\rm b})/0.251 \msol/h$, and then $256^3$ gas resolution elements with an initial mass of $1.27 \times 10^7 \msol/h$. The suite is specialized for training machine learning algorithms and is ideal for this work. The full suite contains different types of runs, of which we use the Latin Hypercube (LH) run. These are 1000 simulations, with varying initial conditions, that span a 6D parameter space of the input parameters noted above. The points in parameter space are sampled via a Latin Hypercube scheme \citep{Heitmann2009Emulator}. The priors for the parameters are $0.1 < \Omega_{\rm m} < 0.5$, $0.6 <  \sigma_8 < 1.0$, $0.25 < A_{\rm SN1, AGN1} < 4$, and $0.5 < A_{\rm SN2, AGN2} < 2$. We also use the dark matter-only (DMO) variant of the LH runs, which are a set of simulations with only dark matter particles and span only $\Om$ and $\sigma_8$.

The \textsc{Camels} simulation suite contains multiple galaxy formation models. In this work, we use the simulations from the galaxy formation model of \textsc{IllustrisTNG}, or TNG, which is a state-of-the-art high-resolution suite of simulations \citep{Nelson2018FirstBimodality, Pillepich2018FirstGalaxies, Springel2018FirstClustering, Marinacci2018FirstFields, Naiman2018FirstEuropium, Nelson2019TNG50, Pillepich2019TNG50}. The model choices made in the fiducial TNG simulations --- which are described in detail in \citet{ Weinberger2017Methods, Pillepich2018Methods} --- are adopted in the \textsc{Camels} simulations. The resolution level of the \textsc{Camels} simulation roughly corresponds to the TNG100-2 run. We will also briefly comment on the \textsc{Astrid} variant of \textsc{Camels}, but present their results in Appendix.

There are four astrophysical parameters varied in the TNG variant of \textsc{Camels}. The total energy injection rate (power) per unit star-formation, $e_w$, and the SN wind speed, $v_w$, are controlled by two normalization factors, $A_{\rm SN1}$ and $A_{\rm SN2}$, respectively. Thus $A_{\rm SN1}$ corresponds to the energy output rate, and $A_{\rm SN2}$ corresponds to the wind speed. The mass loading factor, $\eta$, which corresponds to the amount of mass transferred away through the SNe winds/feedback, is set by the combination $v_w^{-2}e_w$ and so is proportional to the combination $\eta \propto A_{\rm SN1}A_{\rm SN2}^{-2}$.

The SMBH model has different modes of feedback --- a quasar/thermal mode, where the gas in the vicinity of the SMBH is thermally heated, and a radio/kinetic mode, where the gas is given a momentum kick instead. The kinetic mode is turned on when the SMBH accretion has entered a lower accretion rate regime, and is the dominant mechanism for halos at the Milky Way-scale and above \citep[\eg][]{Weinberger2018SMBHsIllustrisTNG, Anbajagane2022BaryImprint}. Note that the kinetic feedback is not continuously output into the halo. Rather, in the TNG model, the SMBH ``accumulates'' energy and expels it once a chosen energy scale is reached, which results in a ``bursty'' feedback process. The parameter $A_{\rm AGN1}$ controls the rate at which energy is accumulated by the SMBH and $A_{\rm AGN2}$ controls the minimum energy needed to initiate such a burst. Both AGN parameters control the rate of the bursts, while the latter also controls the speed of the kinetic feedback jets.

The uber halo sample used in this work --- which is obtained by combining halos from all 1000 LH simulations --- spans a mass range of $10^{11} \msun/h < \Mvir < 10^{14.5} \msun/h$; the minimum mass is set by requiring at least $\approx 1000$ particles in a halo, and the maximum mass is set by the number of available halos. We will still present results up to $\Mvir \sim 10^{15} \msun/h$ as our modeling approach provides uncertainty estimates, which consistently reflect the lack of information at higher mass scales. The uber sample, as well as the mass range we use, follows the choices from \citetalias{\Shao}. Note that, since the maximum mass of a halo in a simulation depends strongly on $\Omega_{\rm m}$, and the value of $\Omega_{\rm m}$ is varied across the 1000 LH simulations, the high mass-regime of the trained model is more heavily informed by simulations with a high $\Omega_{\rm m}$. The maximum halo mass in a single LH sim varies from a minimum of $10^{12}\msol/h$ to a maximum of $10^{14.7} \msol/h$ with a median of $10^{13.7} \msol/h$. This means our model predictions for $\Mvir > 10^{13.7}\msol/h$ are informed by half, or fewer, of the 1000 LH simulations.

\subsection{Normalizing flows}\label{sec:Nflows}

To probe the distribution of halo concentrations, the most direct non-parametric approach would be to estimate the conditional distribution of concentration as a function of all the quantities whose effects we wish to study, including halo mass, and the astrophysical and cosmological parameters. In principle, this could be achieved by constructing a multidimensional histogram of the joint distribution and then conditioning on the relevant halo and simulation properties \citep{Silverman:1986:DensityEstimation,Scott:1992:MultivariateDensity}. However, the number of bins required to maintain a fixed resolution grows exponentially with the dimensionality of the parameter space, an instance of the ``curse of dimensionality'' \citep{Bellman:1961:CurseDimensionality,Scott:1992:MultivariateDensity}. For a finite sample, the available halos are consequently divided among a rapidly increasing number of sparsely populated bins, producing a noisy estimate that can depend sensitively on the chosen bin widths and boundaries \citep{Silverman:1986:DensityEstimation,Scott:1992:MultivariateDensity}. Because the number of available halo samples is limited, directly constructing such a high-dimensional histogram would provide an unreliable representation of the underlying probability distribution. An alternative is to learn a smooth conditional-density model whose parameters vary continuously with the conditioning variables, allowing information to be shared across neighboring regions of parameter space \citep{Bishop:1994:MDN,Papamakarios:2016:ConditionalDensity,Papamakarios:2017:MAF}. Neural density-estimation (NDE) methods have consequently become important tools for modeling complex conditional distributions and for simulation-based inference \citep{Cranmer:2020:SBI,Alsing:2018:DELFI,Alsing:2019:pydelfi}.

Normalizing flows (NFs) are one such NDE method that has been widely adopted for learning probability distributions in cosmology \citep{Alsing:2018:DELFI,Alsing:2019:pydelfi,Jeffrey:2021:DESLFI}. A NF models a complex target distribution by transforming a simple base distribution, such as a multivariate normal distribution, through a sequence of invertible and differentiable mappings \citep{Rezende:2015:NormalizingFlows,Papamakarios:2017:MAF,Papamakarios:2021:NormalizingFlows}. These mappings must satisfy these conditions of being invertible and differentiable so that the transformation is reversible between the target and base distributions. 
Let $Y$ denote a random variable following the target distribution and $Z$ denote a random variable drawn from the known base distribution, which is Gaussian in our work. If $g$ represents the sequence of transformations from the base distribution to the target distribution, then $Y=g(Z)$ and $Z=f(Y)$, where $f=g^{-1}$ is the inverse transformation that maps the target distribution back to the base distribution, or the ``normalizing'' direction. This reversibility allows the likelihood of the observed data to be evaluated using the change-of-variables formula with Jacobian $J$:
\begin{align}
    p_Y(y)
    &= p_Y(g(z)) \\
    &= p_Z(z)\left|\det J_g(z)\right|^{-1} \\
    &= p_Z(f(y))\left|\det J_f(y)\right|.
\end{align}
If $g$ and $f$ each represent a sequence of $N$ transformations,
\begin{align}
    g &= g_N \circ g_{N-1} \circ ... \circ g_1, \\
    f &= f_1 \circ f_2 \circ ... \circ f_N,
\end{align}
then the Jacobian determinant of the full inverse transformation is given by
\begin{align}
    \left|\det J_f(y)\right|
    &= \prod_{i=1}^{N}
    \left|\det J_{f_i}(u_{i-1})\right|,
\end{align}
where $u_{i-1}$ is the input to the $i^{\rm th}$ inverse transformation. Equivalently, the logarithm of the target density is
\begin{align}
    \ln p_Y(y)
    &= \ln p_Z(f(y))
    + \sum_{i=1}^{N}
    \ln\left|\det J_{f_i}(u_{i-1})\right|.
\end{align}
The transformations $g$ and $f$ may additionally depend on a set of conditioning variables, which in our case are the halo mass and the six cosmological and astrophysical simulation parameters. We therefore model the conditional probability distribution
\begin{align}
    p_Y(y\mid\boldsymbol{\theta})
    &= p_Z(f(y;\boldsymbol{\theta}))
    \left|\det\frac{\partial f(y;\boldsymbol{\theta})}{\partial y}\right|,
\end{align}
where $\boldsymbol{\theta}$ denotes the seven conditioning variables.

Our measurement of interest are the moments of the distribution, $P(\ln \cvir | \ln \Mvir)$. The mean is computed trivially, 
\begin{equation}\label{eqn:mean}
    \langle \cvir \ln \cvir | \ln \Mvir \rangle  = \int d\ln \cvir \,\,P(\ln \cvir | \ln \Mvir )
\end{equation}
while the higher-order moments are calculated as,
\begin{align}\label{eqn:moments}
    m^N = \int d\ln \cvir \bigg[&(\ln\cvir - \langle \ln \cvir | \ln \Mvir \rangle)^N \nonumber\\
    & \times P(\ln \cvir | \ln \Mvir )\bigg]
\end{align}
where all terms in the above two expressions are also implicitly conditioned on the vector, $\boldsymbol{\theta}$. We denote the skewness as $\gamma \equiv m^3 / \langle m^2 \rangle^{3/2}$ and the \textit{excess} kurtosis, $\kappa = \langle m^4\rangle / \langle m^2 \rangle ^2 - 3$. For brevity, we will refer to it as kurtosis. Given the above definitions, $\kappa$ obeys the inequality, $\kappa \geq \gamma^2 - 2$. We will also present the standard deviation rather than the $m^2$ (variance) moment, and present it as $\sigma(\ln \cvir)$ and not $\sigma (\log_{10} \cvir)$ --- which are connected by a constant factor, $\ln(10)$ --- as the former is easily interpreted as a fractional scatter on $\ln \cvir$.

To estimate the uncertainty in the distribution modeled by the NF (and its derived moments), we re-trained the NF models on 100 bootstrap-resampled realizations of the user sample in our simulations. This provides us with 100 NF models from which we can derive any statistics of interest. These 100 realizations will also include any stochastic variations from the NF training as well. The results discussed in this work quote uncertainties obtained from this above procedure. We find the modeling uncertainties mimic expected behaviors, i.e., they increase towards higher halo masses where the effective sample is dramatically smaller than that at lower masses.

In this work, we use the NDEs implemented in \textsc{PyDelfi} \citep{Alsing:2019:pydelfi} to model the conditional distribution of halo concentration. Our estimator ensemble consists of one conditional masked autoregressive flow \citep[MAF,][]{Papamakarios:2017:MAF}, constructed from a sequence of five MADE transformations \citep{Germain:2015:MADE}, together with five independently trained mixture density networks \citep[MDNs,][]{Bishop:1994:MDN}. The five MDNs employ between one and five Gaussian mixture components, respectively, allowing progressively more flexible descriptions of non-Gaussian features in the conditional distribution. The parameters of the MDNs are predicted by neural networks which take the halo mass and the six simulation parameter as input. The probability densities predicted by the six NDEs above are combined by the \textsc{PyDelfi} weighted stacking procedure, where the weights are the validation losses of each model. In our case, the weighting in the stacking is such that the NF dominates the final predictions. Hence, our discussion below will only refer to the NF.

\section{Dependence on cosmology in dark matter-only models}\label{sec:DMO}

\begin{figure}
    \centering
    \includegraphics[width=1\columnwidth]{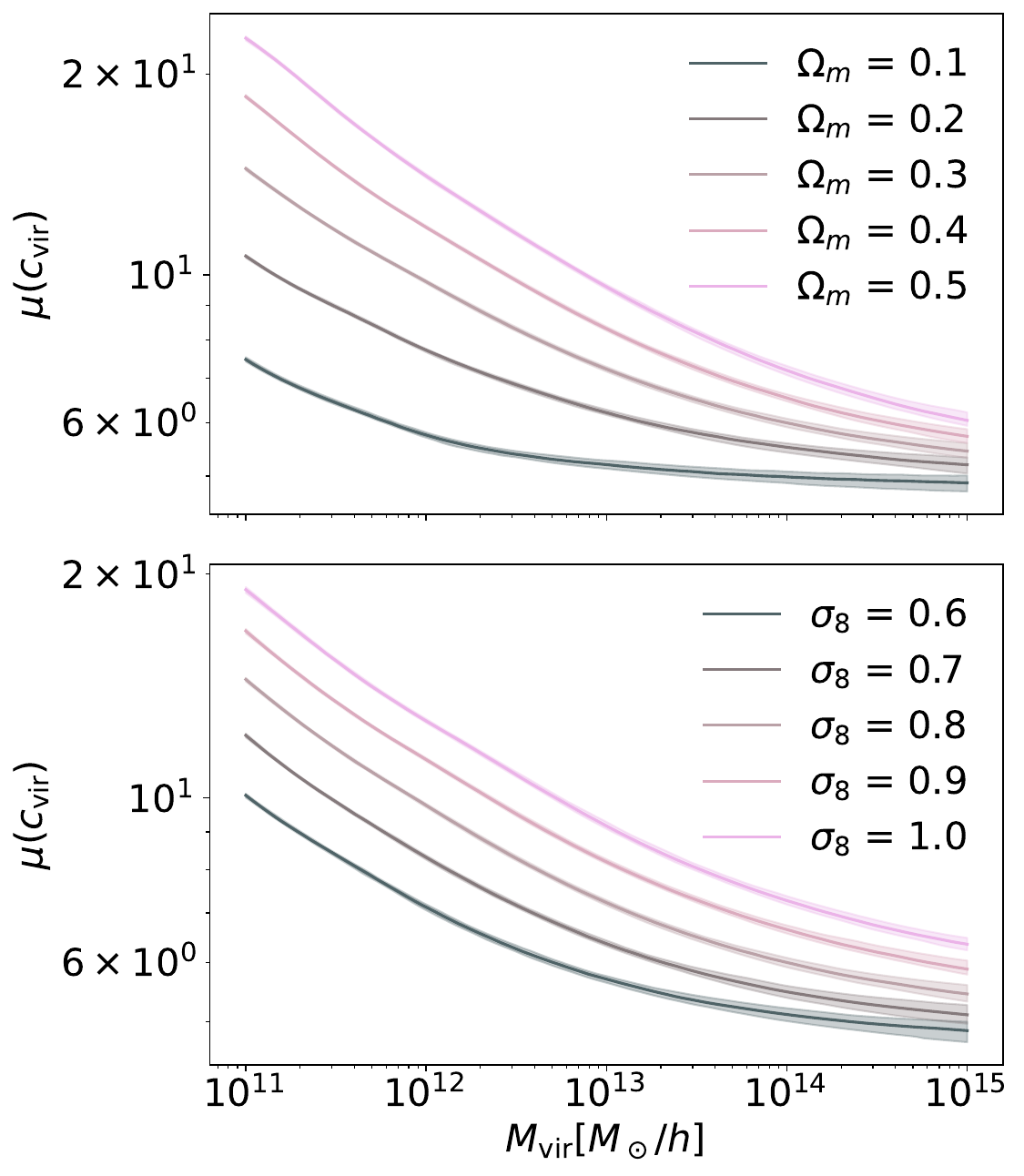}
    \caption{The first moment (mean) of the concentration--mass relation as a function of $\Omega_{\rm m}$ (top) and $\sigma_8$ (bottom). The predictions are obtained from our NF model, trained on the \textsc{Camels} simulations. Uncertainties are obtained by training 100 different flow-based models on 100 bootstrap-resampled datasets.}
    \label{fig:dmo_mean}
\end{figure}

\begin{figure*}
    \centering
    \includegraphics[width=1\linewidth]{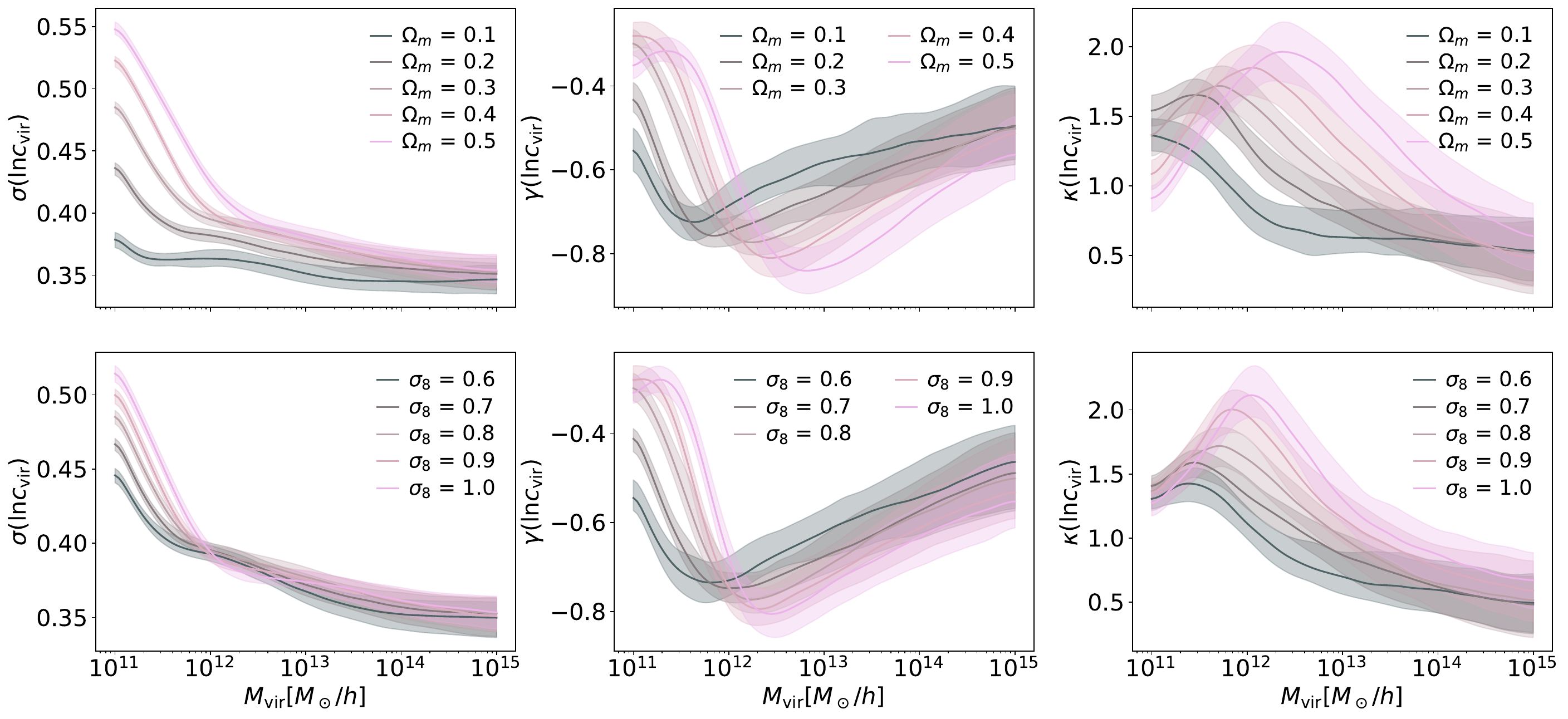}
    \caption{The second to fourth moments of the concentration--mass relation as a function of $\Omega_{\rm m}$ (top) and $\sigma_8$ (bottom). The scatter is presented as $\sigma (\ln \cvir)$, not $\sigma (\log_{10}\cvir)$ as the former can more easily be interpreted as a fraction uncertainty.}
    \label{fig:dmo_HOM}
\end{figure*}

We start by discussing the dependence of the concentration distribution on cosmology in a DMO simulation where matter interacts only gravitationally. This sets the baseline expectation for the evolution of the moments, and gives us a reference to compare against when discussing the behavior in simulations with galaxy formation (Section \ref{sec:hydro}).

Figure \ref{fig:dmo_mean} shows the mean relation across the full mass range, as a function of $\Omega_{\rm m}$ and $\sigma_8$. This relation exhibits a dependence on cosmology that follows the well-established findings from \citet{Diemer2019concentrations}.\footnote{While we do not expect precise agreement with \citet{Diemer2019concentrations} given the significant differences between the simulations used in that work and the \textsc{Camels} simulations we use here, we have separately confirmed that we produce a close match to their prediction for the mean concentration and that our model exhibits the same mass-dependent change with cosmology as their work.} The direction of the change is also as anticipated: $\Om$ and $\seight$ both control the matter distribution, and the ease with which structure can form. The concentration is well known to be tied directly to the accretion rate of halos \citep{Wechsler2002Concentrations, Dalal2010OriginsDmHaloProfiles}, which in turn is tied to the halo formation time. Increasing $\Om$ and $\seight$ pushes the halo formation to earlier times and results in more concentrated halos across all masses.

Moving onto the second moment, Figure \ref{fig:dmo_HOM} shows the scatter is largely constant at $\sigma(\ln \cvir) \approx 0.35$,\footnote{We always report the scatter on the natural log of the concentration, $\ln \cvir$, as this can be easily interpreted as a fractional scatter.} but increases towards lower masses. \citet[][henceforth, \citetalias{\Anba}]{Anbajagane2022BaryImprint} found a similar increase, and linked it to the presence of tidal interactions between these lower mass halos and their more massive neighbors; these interactions generally strip the halo of its outer mass distribution and cause it to appear highly concentrated as a result \citep[][see their Figure 6]{Mansfield2020AssemblyBias}. Section 4.2 in \citetalias{\Anba} also notes that removing such tidally affected low-mass objects reduces the increase in scatter. The $\Om$ dependence of this low-mass scatter is also expected, as a higher $\Om$ value increases the number density of halos in the simulations and therefore, increases the likelihood for such tidal interactions to occur. Note that for larger halos, $\Mvir > 10^{13} \msun/h$, there is little-to-no change in the scatter. This is consistent with \citet{Kwan:2013:cM}, who found the scatter in this mass regime is broadly constant (see their Figure 2) across 37 cosmological models.

Moving onto the higher-order moments, we find the skewness is always negative, indicating the presence of a tail to low concentration values. This has been identified in previous works, and is caused by unrelaxed halos \citep[\eg][see their Figure 4]{Maccio2008CosmoDep}, often caused by recent mergers, which disturb the halo core and cause it to become more diffuse. Note that this is distinct from the tidal-interaction event noted above, and will decrease the halo concentration. The mass-dependence of the skewness is non-monotonic and shows a clear maximum at around Milky Way (MW)-mass scales. However, this mass scale shifts both with $\Om$ and with $\sigma_8$. In all cases, the mass scale here coincides with the mass scale where the scatter estimate sharply increases. One may naively expect this to occur due to our definition of the skewness --- if the third moment of the distribution is fixed, then for larger scatter values the skewness will reduce. However, we have confirmed that the skewness also has a non-monotonic behavior with mass. Instead, the coinciding mass scales suggest that this change in skewness at low masses is caused by the same tidal stripping effect that we discussed before. Tidal interactions generate a tail to high concentration values \citep{Mansfield2020AssemblyBias}, which offsets the tail to low values that is caused by mergers.

The kurtosis also has a mild-dependence on cosmology, though it largely mimics that of the skewness. The exact equality, $\kappa = \gamma^2 - 2$ is not satisfied, indicating there are additional contributions to the kurtosis than just the skewness alone. Though, such contributions are subdominant relative to the skewness contributions. Thus, the weight in the tails of the distribution (which is captured by the kurtosis) is largely informed by just the skewness of the distribution. The exception is for MW-mass systems, where changes to $\seight$ alter the kurtosis by a factor of 2, while they do not affect the skewness at all. Objects at these masses are resolved by 1000 particles, so they are considered well-resolved, though they will still incur some resolution effects \citep{Mansfield2021SimBias}. We do not expect these effects to complete destroy a physical signature, as the same resolution-based bias will be present in all the simulations we use, and so relative changes between the simulations will be significantly less impacted by such resolution effects. We have previously noted that the low mass regime is impacted by a tail to both high and low concentrations caused by  tidal interactions and mergers, respectively. This may suggest that the kurtosis should be increasing for small halos. However, both interactions are also known to alter the scatter of the halo concentration \citep{Wechsler2002Concentrations, Duffy2008Concentration, Mansfield2020AssemblyBias, Anbajagane2022BaryImprint}. Thus, the tails of the distribution become less heavy because the bulk of the distribution widens as well.

The above picture is largely consistent with the existing understanding of the distribution $P(\ln \cvir)$ --- it is generally non-Gaussian due to interactions between halos \citep{Neto2007Concentration,Maccio2008CosmoDep,Lee2018TidalStripping} but returns to a Gaussian form once unrelaxed and tidally stripped halos are removed from the analysis \citep{Jing2000Concentrations,Neto2007Concentration}. Such unrelaxed (tidally stripped) halos form a skewness towards low (high) concentrations, and will naturally cause both (i) an increase in the estimated scatter, and (ii) a kurtosis feature. The cosmology dependence of the scatter, skewness, and kurtosis can therefore be interpreted through the impact of cosmology on halo mass-accretion histories and merger statistics \citep{Wechsler2002Concentrations,Ludlow2014MilleniumConcentration,Correa2015Concentration,Fakhouri2010ModelMAH}, which in turn determine the frequency and evolutionary stage of recent mergers and tidal-processing events \citep{Lee2018TidalStripping,Mansfield2020AssemblyBias}.

\section{Dependence on cosmology and astrophysics in hydrodynamical models}\label{sec:hydro}

\begin{figure*}
    \centering
    \includegraphics[width=1\linewidth]{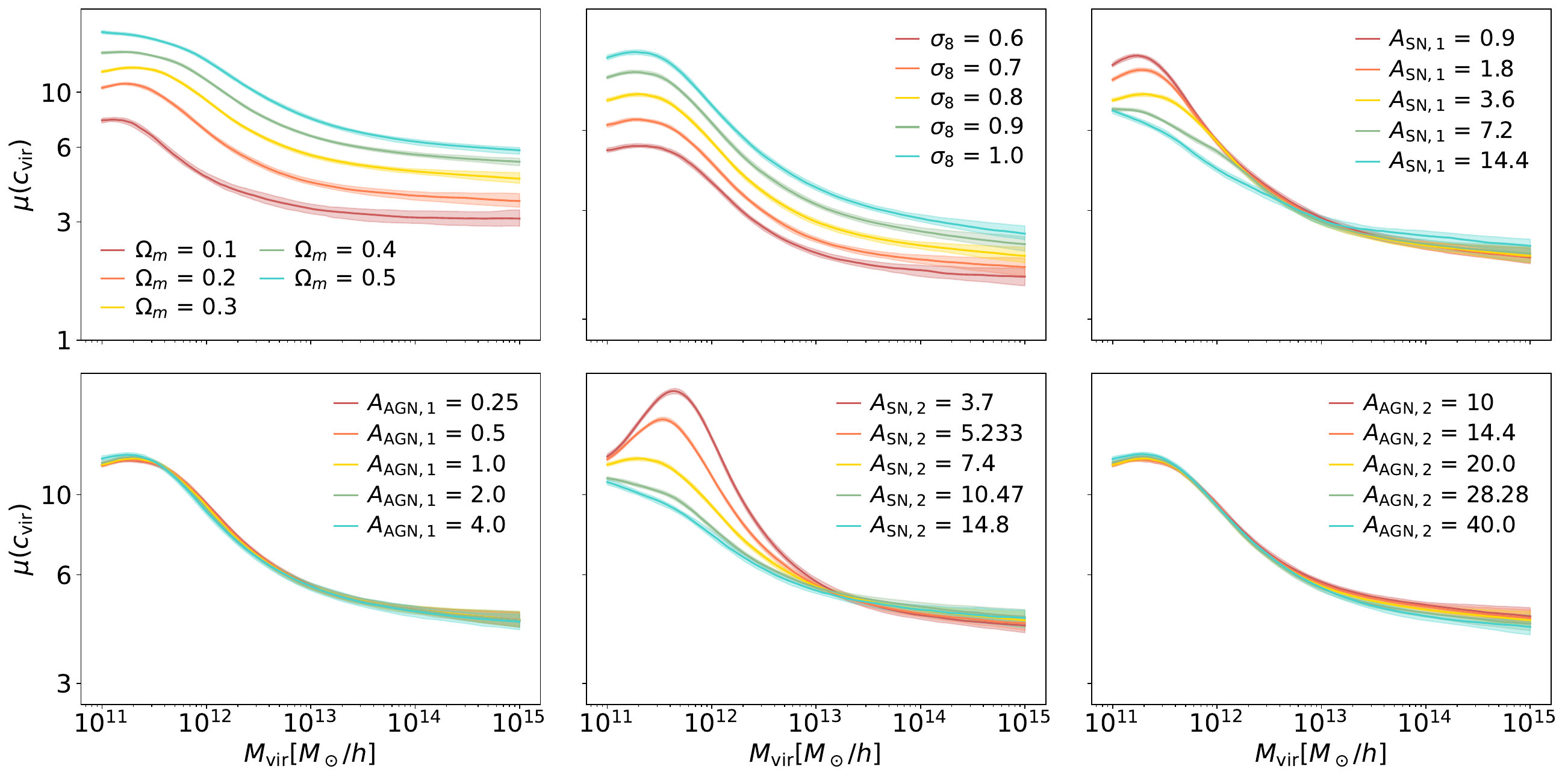}
    \caption{The mean relation, $\langle \ln \cvir | \ln \Mvir\rangle$, as a function of the six simulation parameters in the TNG run of \textsc{Camels}. There are clear dependences on $\Om$, $\seight$, $A_{\rm SN1}$, and $A_{\rm SN2}$. There is a mild-dependence on $A_{\rm AGN2}$ in the high-mass regime. All features are consistent with existing analyses of halo concentration in the \textsc{Camels} TNG model.}
    \label{fig:mean}
\end{figure*}

We now extract the distribution $P(\ln \cvir | \ln \Mvir)$ in the \textsc{Camels} simulations with galaxy formation physics. We split our discussion separately across the four different moments that we measure. While we focus on the TNG model for our main results, Appendix \ref{appx:astrid} also shows the behavior from the ASTRID model, which is qualitatively similar to the behavior we discuss further below. The interested reader can look at Appendix \ref{appx:astrid} for more details.

\subsection{Mean}\label{sec:hydro:mean}

Our estimate of the mean relation, $\langle \ln \cvir | \ln \Mvir\rangle$, must follow that of \citetalias{\Shao}, who used the same \textsc{Camels} simulations we employ here. Indeed, we find that our Figure \ref{fig:mean} reproduces the behavior of Figure 6 in \citetalias{\Shao}. In line with the latter's result, the mean relation is most sensitive to $\Om$ and $\seight$, with lower-mass halos also showing a clear sensitivity to $A_{\rm SN1}$ and $A_{\rm SN2}$. The higher-mass halos have a mild sensitivity to the $A_{\rm AGN2}$ parameter as well. As noted, all these features are qualitatively consistent with the results of \citetalias{\Shao}, who used a local linear regression estimator for their analysis. This comparison provides more confidence, in addition to the validations in Appendix \ref{appx:linregress}, that our NF approach is providing qualitatively and quantitatively reasonable results.

For completeness, we briefly discuss the phenomenology connecting the six simulation parameters and the mean relation, but direct readers to \citetalias{\Shao} for more details. The connection to cosmological parameters largely follows from the discussion in the DMO case above. The only difference is that $\Om$ now also controls the baryon fraction, $f_b = \Omega_{\rm b}/\Om$, as the simulations are all produced at fixed $\Omega_{\rm b} = 0.049$, and varying this fraction will alter the impact of astrophysics on the halo. The astrophysical parameters behave as per their naive interpretation; increasing their values makes feedback stronger and leads to less concentrated halos. However, we also demonstrate the non-linear coupling discussed in \citet{Shao2023Baryons} and \citet{Gebhardt2023CamelsAGNSN}, where increasing SNe feedback ($A_{\rm SN2}$) \textit{increases} the halo concentration in massive halos. This is statistically significant, and is a result of non-linear coupling of processes --- strong SNe feedback disperses gas away from the central region of the halo, i.e., away from the AGN, and therefore reduces the impact of AGN feedback. This, in turn, enables the halo to be more concentrated. Such a phenomenology arises because SNe feedback is strong enough to disperse matter within a halo, but often not energetic enough to eject it outside the halo (especially for larger halos with deeper gravitational potentials). Whereas, the AGN feedback can easily eject matter outside the halo virial radius. See \citet{Gebhardt2023CamelsAGNSN} for more details on the ejection scale observed in the \textsc{Camels} simulations.

\subsection{Scatter} \label{sec:hydro:std}

\begin{figure*}
    \centering
    \includegraphics[width=1\linewidth]{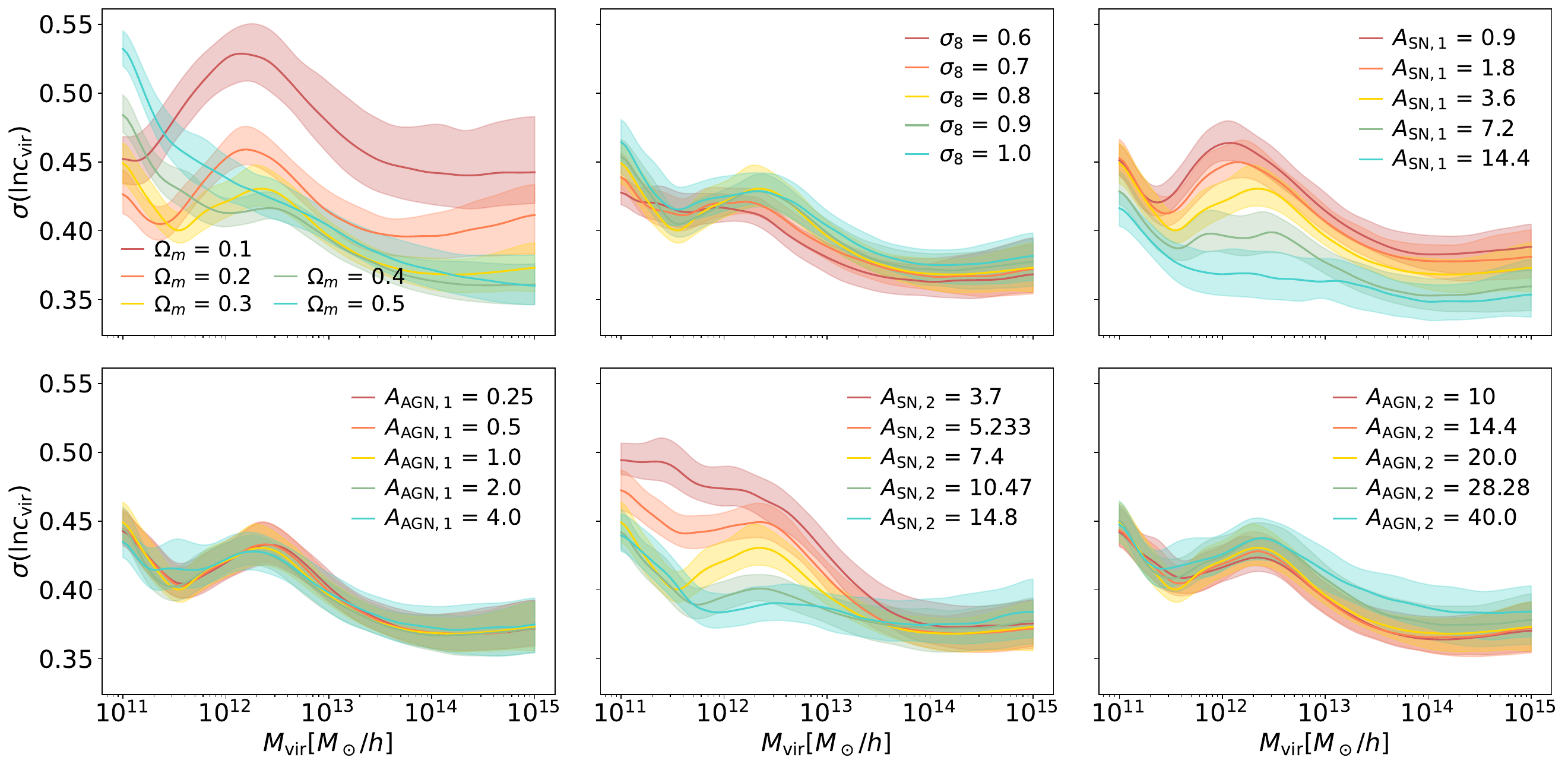}
    \caption{Similar to Figure \ref{fig:mean}, but for the second moment (scatter) of the distribution. We show $\sigma(\ln \cvir)$, which can be interpreted as a fractional scatter on $\cvir$.}
    \label{fig:std}
\end{figure*}

Figure \ref{fig:std} focuses on the scatter of the concentration distribution. There is a minor-to-no dependence on the parameters $\seight$, $A_{\rm AGN1}$, and $A_{\rm AGN2}$. The scatter shows a clear increase towards low $\Om$. However, when restricting to $\Om \in [0.25,0.4]$, which is a more observationally viable range  \citep[\eg][]{Planck2020CosmoParams,AbdulKarim:2025:DESIDR2BAO,Abbott:2024:DESSN5YR,DES2022Y3,Asgari2021KidsWL,Anbajagane2025d,Anbajagane2025e}, the scatter at higher masses is consistent with no statistically significant dependence on $\Om$. At lower masses, the impact of unrelaxed halos is once again important. As noted previously, the dependence on $\Om$ is better understood through the baryon fraction. At low $\Om$, the ratio is $f_b \sim 0.5$ (a three-fold increase compared to the fiducial, $f_b \approx 0.15$) and results in baryons becoming a more-dominant part of the $\cvir$ estimate. Thus, the scatter in the gas/star distributions --- which can be larger than that of the dark matter (DM) distribution given the variety of non-gravitational processes affecting the former --- dominates the $\cvir$ estimate and results in a larger scatter. This also explains why increasing $f_b$ induces a peak in the scatter at $M \approx 10^{12} \msun/h$. The latter is the characteristic mass scale of peak star-formation efficiency \citep{Wechsler2018GalaxyHaloConnection}, and increasing the available reservoir of baryons (relative to DM) increases the impact of star formation at these halo-mass scales.

The two SNe parameters have a notable impact on the scatter as well. Following the arguments above, their impact scales with how they affect star formation efficiency. For lower values of the parameters, there is less feedback and therefore more gas cooling and star formation. Lower values of $A_{\rm SN1}$ (energy per SN) exhibits the prominent peak we discussed previously, while the same for $A_{\rm SN2}$ (wind speed) also increases the scatter at even smaller halo masses. This is reasonable, as the wind speed (which, at fixed energy rates, is coupled to the mass loading factor) can greatly stifle star-formation rates in smaller halos in particular.

We also note the somewhat surprising lack of any impact from changing $\seight$. In Section \ref{sec:DMO}, we noted that this parameter impacts the halo formation time/history, which is known to be one of the dominant drivers of scatter in halo properties and particularly in the halo concentration. However, in the hydrodynamical case, the astrophysical processes appear to dominate the scatter instead. The relevance of recent mergers and tidal interactions\footnote{We note that a compact central stellar component can reduce a smaller halo's susceptibility to tidal stripping and disruption by a larger object, as the stellar component increases its central density and deepens its gravitational potential well \citep{Nagai:2005:RadialDistribution,Weinberg:2008:BaryonDynamics,Dolag2009Subfind,Zhu2016ConcentrationIllustris}.}
 appears to be minimized as there is no longer a significant increase in scatter towards lower-mass halos as seen in Figure \ref{fig:dmo_mean}. The amplification re-emerges if we increase $\Om$, thereby returning the simulation closer to a DMO case. This provides further evidence that the evolution of the gas/star components is suppressing the tidal disruption-based increase in the scatter.

\subsection{Skewness}\label{sec:hydro:skewness}

\begin{figure*}
    \centering
    \includegraphics[width=1\linewidth]{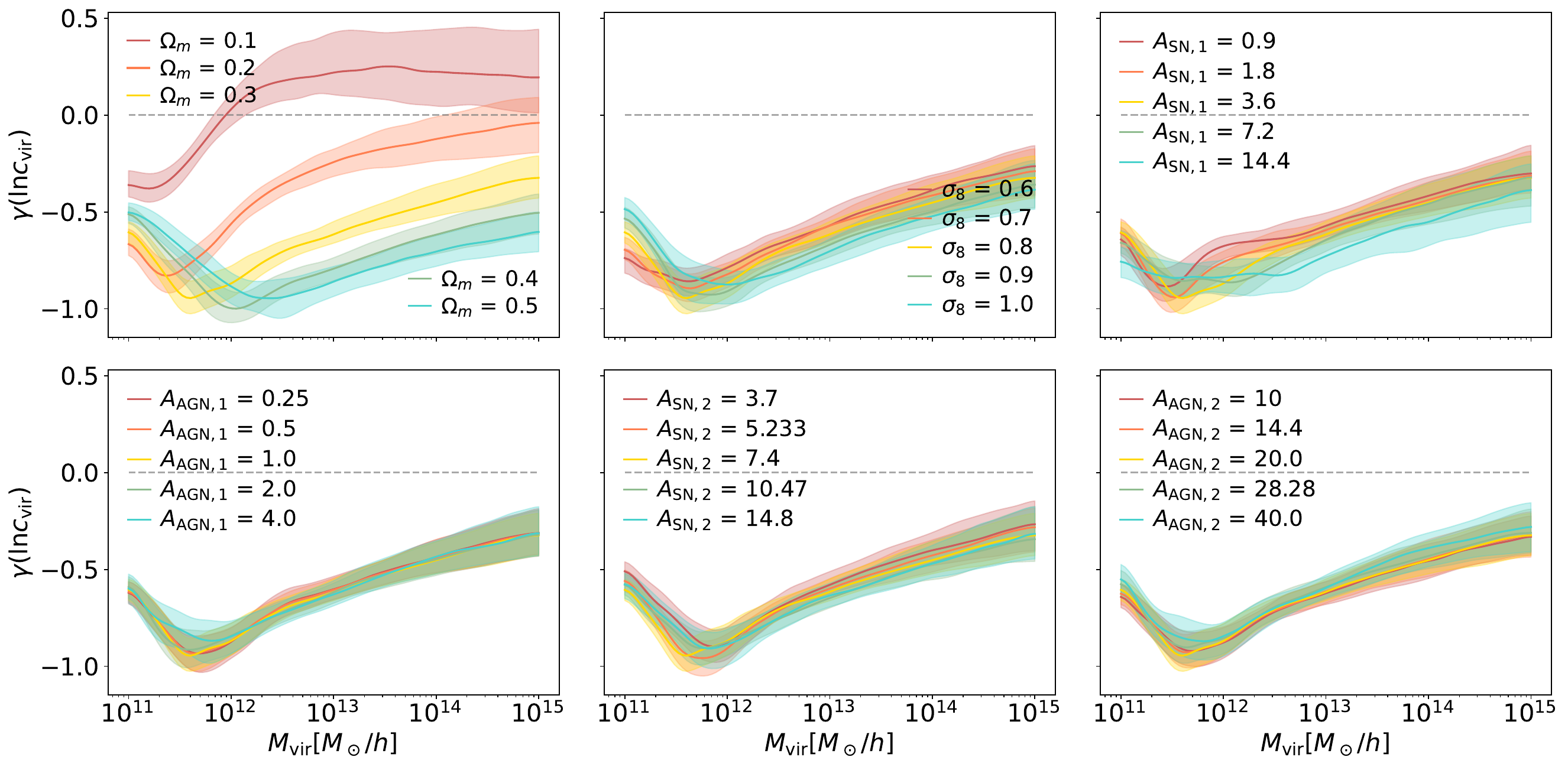}
    \caption{Similar to Figure \ref{fig:mean}, but for the skewness of the concentration--mass distribution.}
    \label{fig:skewness}
\end{figure*}

The skewness for the fiducial choice of parameters has a fairly similar shape to that in the DMO scenario, and the amplitudes are quite similar. The skewness is weaker (stronger) at large (small) halo masses. It is also always negative, showing that the asymmetric tail is towards low concentration values. These can generally be caused by merger events, as in the DMO case, but also AGN feedback events that eject material well outside the halo. We are unable to distinguish the relative importance of these two without detailed extraction of AGN energy information from \textsc{Camels}. Given the similar mass-dependent shape of the skewness measurement here compared to that of the DMO case, we expect that the merger events (which are the only factor in the DMO case) are the more dominant effect of the two.

The skewness is also very stable to changes in the astrophysical parameters. There is a minor dependence on $\seight$, which is consistent with our view above that the skewness is predominantly informed by merger rates, and indeed larger values of $\seight$ (which promote more halo growth, and therefore merger events) increases the skewness. The impact from $\Om$ is an order of magnitude larger. Unlike for the DMO case, the skewness now reduces when decreasing $\Om$. In practice, this originates from the same behavior discussed before -- for low $\Om$, the stellar distribution is more dominant to the total matter distribution fits. This then affects the tails of the $\cvir$ distribution. The preference for a positive tail is understandable, as a higher stellar fraction means a larger fraction of the halo is made up of the centrally peaked stellar component. This will increase the halo concentration measured for a given halo. There is also a hint of a dependence on $A_{\rm SN,1}$, but only at MW mass scales and with lower values of the parameter reducing the skewness. Such a feature would be consistent with a stellar-mass based origin, as weakened SNe encourage more star formation and therefore let the stellar core dominate the mass distribution. As a result, we have more high concentration values, which slightly suppresses the fiducial, negative skewness.

In summary, the skewness in the concentration largely follows the behavior found in DMO simulations. The most striking feature is the increased dependence on $\Om$, where lower values allow the stellar distribution to dominate the halo more and create a tail towards high concentration values. The dependence on the remaining five parameters is either mild or undetected.

\subsection{Kurtosis} \label{sec:hydro:kurtosis}

\begin{figure*}
    \centering
    \includegraphics[width=1\linewidth]{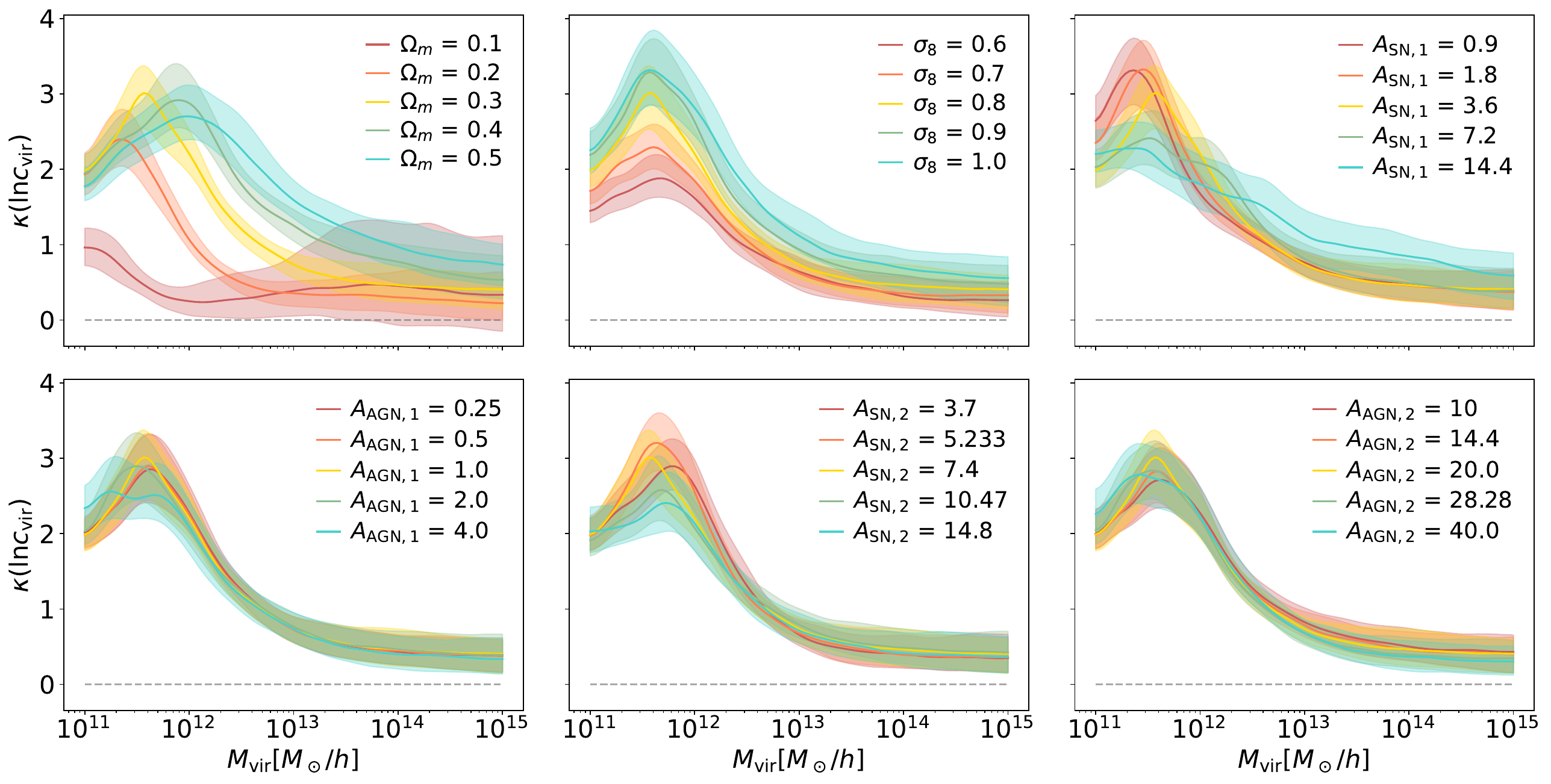}
    \caption{Similar to Figure \ref{fig:mean} but for the excess kurtosis of the concentration--mass distribution.}
    \label{fig:kurtosis}
\end{figure*}

Finally, we discuss the excess kurtosis of the concentration distribution, as presented in Figure \ref{fig:kurtosis}. Once again, the distribution is largely insensitive to the astrophysical parameters being varied. However, there is a clear difference relative to the DMO case in Figure \ref{fig:dmo_mean}; the excess kurtosis in the DMO case peaks at $\kappa = 1.6$ whereas in the hydrodynamical case it peaks at $\kappa = 2.5$ instead. Thus, it appears galaxy formation does impact the kurtosis but the four parameters being varied here do not have a particularly large impact on the emergent behavior. 

The overall trend of the excess kurtosis with the cosmology parameters can be expected given the skewness findings in Figure \ref{fig:skewness} and given the consistency relation, $\kappa \geq \gamma^2 - 2$. Since $\gamma$ reduces for lower values of $\Om$, we expect $\kappa$ will likely reduce as well. The peaks of the non-monotonic mass-dependence shift when varying $\Om$ and their mass-scale coincides with same for the peaks in the skewness, as shown in Figure \ref{fig:skewness}. Similarly, the trend of $\kappa$ with $\seight$ follows the expected direction and amplitude based on the trend of $\gamma$ with $\seight$. Overall, the kurtosis does not show any behaviors that differ significantly from the expectations set by the behavior of the skewness.

\section{Conclusion}\label{sec:conclusion}

The halo model is a ubiquitous tool for modeling the cosmological structure in our Universe, spanning kiloparsecs to gigaparsecs. The halo concentration is a key ingredient in this model, particular on length scales smaller than a few megaparsecs (i.e. the largest halo scales). Most analyses only consider the mean concentration at fixed halo mass, while some also include the scatter in this relation. Other analyses, such as the modeling of strong lenses, depend on the full distribution of the halo concentration and model tend to model it as a Gaussian in log concentration with a fixed cosmology-independent scatter. In this work, we quantify the higher moments of the concentration--mass relation using a normalizing flow model and extract the moments' dependence on both cosmology and astrophysics. We study the connection with cosmology in a dark matter-only (DMO) simulation, as a reference, and then extend to simulations with the IllustrisTNG galaxy formation model.

The main results from our work are as follows:
\begin{itemize}
    \item In DMO simulations, the scatter in halo concentration is nearly independent of cosmology for higher-mass objects but increases by a factor of 1.6 form simulations with higher $\Om$ and $\seight$ (Figure \ref{fig:dmo_HOM}). This is connected directly to the increased presence of active mergers and tidally stripped, low-mass systems.

    \item The higher moments of the concentration in these DMO simulations are non-zero and show a clear dependence on halo mass and on cosmology (Figure \ref{fig:dmo_HOM}). At higher masses, the concentration is skewed low by active mergers causing more diffuse halos, while at lower masses, tidal stripping results in more concentrated objects.

    \item In simulations with galaxy formation, the mean relation shows a dependence on cosmology and astrophysics that is consistent with existing \textsc{Camels}-based work (Figure \ref{fig:mean}). The scatter has a significant dependence on the baryon fraction and on the SNe feedback strength (Figure \ref{fig:std}), as increasing the baryon fraction causes the concentration scatter to be dominated by astrophysical processes rather than by mergers and tidal stripping. The astrophysical effects result in a non-monotonic increase in the scatter at Milky Way-mass scale.

    \item The skewness in the concentration, when the simulations include galaxy formation, is qualitatively similar to that that measured in DMO simulations. It shows a large sensitivity to the baryon fraction but is insensitive to the other parameters (Figure \ref{fig:skewness}). 
    
    \item The behavior in the kurtosis follows largely from that of the skewness (Figure \ref{fig:kurtosis}) as the two are linked by an inequality condition.
\end{itemize}

In general, we show that the distribution in halo concentration (after subtracting out the mean relation) is sensitive to cosmology in the DMO case, and to both cosmology and astrophysics when including galaxy formation physics. The skewness and kurtosis of this distribution is non-zero in all cases, with the non-Gaussian nature of the PDF being dominated by the skewness. The importance of this behavior in models that use the halo concentration depends entirely on the chosen model and its sensitivity to the mean concentration compared to the full distribution of the concentration.

Throughout this work, we have focused on the IllustrisTNG model of galaxy formation. However, the properties of baryons in halos vary when considering simulations with different galaxy formation models \citep[\eg][]{Farahi2018BAHAMAS, Chua2019ShapeIllustrisBaryons, Anbajagane2020StellarStatistics, Chua2021TNGShapes, Lim2021GasMass, Anbajagane2022GalVelBias, Stiskalek2022ML}. The distribution of halo concentration can therefore have sensitivity to the exact underlying model. We have briefly checked that the behavior discussed above is not qualitatively different when considering one another model provided in the \textsc{Camels} suite, based on the ASTRID galaxy formation prescription. However, future work will need to study these simulation variants in more detail to generalize our results. Given the behavior in the skewness mimics that of the DMO case, we hypothesize that its behavior is fairly general and will be present across different galaxy formation models. 

As analyses of the ultra small dark-matter halos become increasingly prevalent, for example in studies of strong lensing systems or stellar streams, the importance of the full distribution of the halo concentration will only grow. Our work presents a useful characterization of the relevant dependencies of such a distribution, and also a robust methodology for characterizing these features. Future efforts will need to utilize high-resolution, large-box simulations to revisit these behaviors (even for the DMO case) and build a model with adequate precise for use in cosmological analyses of these rich and varied observable systems.

\section*{Acknowledgements}

We thank the \textsc{Camels} collaboration for making their data publicly available and easily accessible to the community. DA is supported by NSF grant No. 2108168.

All analysis in this work was enabled greatly by the following software: \textsc{Pandas} \citep{Mckinney2011pandas}, \textsc{NumPy} \citep{vanderWalt2011Numpy}, \textsc{SciPy} \citep{Virtanen2020Scipy}, and \textsc{Matplotlib} \citep{Hunter2007Matplotlib}. We have also used
the Astrophysics Data Service (\href{https://ui.adsabs.harvard.edu/}{ADS}) and \href{https://arxiv.org/}{\texttt{arXiv}} preprint repository extensively during this project and the writing of the paper.

\section*{Data Availability}

The public \textsc{Camels} data, including the \textsc{Rockstar} catalogs we use in this work, can be found at \url{https://camels.readthedocs.io}.

\bibliographystyle{mnras}
\bibliography{References}

\appendix

\section{Validation of normalizing flows}\label{appx:linregress}

\begin{figure*}
    \centering
    \includegraphics[width=1\linewidth]{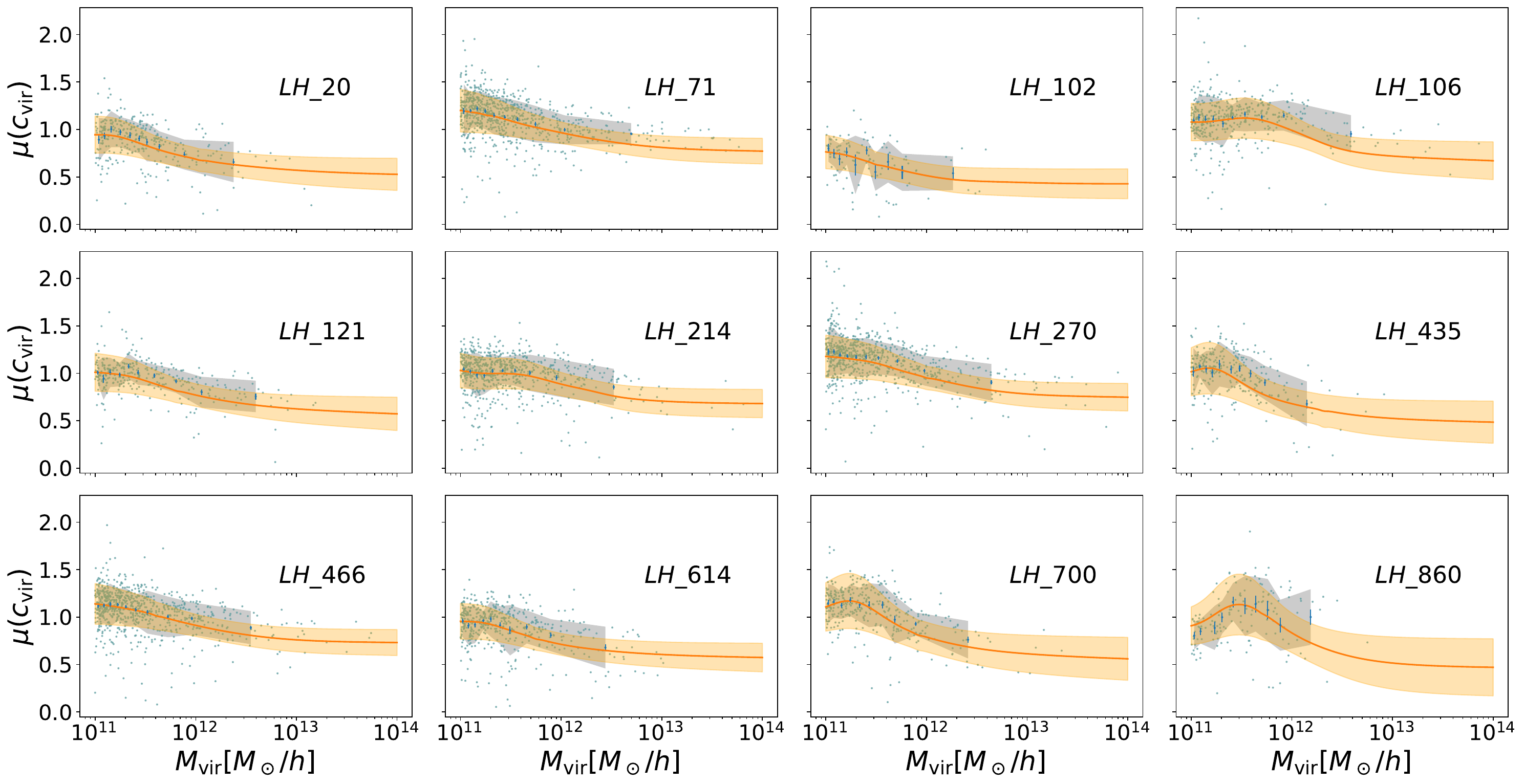}
    \caption{Comparison of the predictions from the normalizing flow to simple mean/variance estimators. We perform this test for twelve individual simulations from the LH suite. The blue points with errorbars are the estimates of the mean from the simulation, with bootstrap-based uncertainties. The standard deviation is shown as the gray band. The predictions for the normalizing flow are shown as the orange line and band. The faint blue points show the full sample in a given simulation. We extend the simulation measurement only up to an $\Mvir$ set by the 20th most massive halo in the given simulation. The NF predictions span the full mass range.}
    \label{fig:validation}
\end{figure*}

Throughout this work, we utilize the NF methodology for estimating the relation between $\cvir$ and $\Mvir$ (and the associated simulation parameters). While the formalism is well established, we must still validate its performance. Here, cross check that the flow-based results are consistent with simple mean and variance estimators.

To validate the mean concentration–mass relation predicted by the NF (Equation \ref{eqn:mean}), we directly measured the $c_{\rm vir}$ distribution across different $M_{\rm vir}$ ranges using the individual LH simulations. We randomly select 12 simulations and, within each simulation, divide the halos into 10 percentile-based bins according to their $M_{\rm vir}$ values. For each bin, we calculate the mean $c_{\rm vir}$ of the halos (blue scatter points with uncertainties) and overlaid the mean $c_{\rm vir}$ predicted by the NF (orange line). The uncertainties on the mean are estimated from bootstrap resampling the individual simulation's catalog. To validate the scatter predictions, we calculate the standard deviation of the $c_{\rm vir}$ distribution within each halo mass bin. This is shown as the gray shaded region. For comparison, we overlay the corresponding 68\% region of the NF predictions as an orange shaded region. 

Both the mean and the scatter predicted by the NFs show overall agreement with the directly measured quantities across all twelve test simulations. This, in conjunction with two aforementioned consistency checks --- (i) agreement of our mean relation in the DMO case with \citet{Diemer2015Concentration}, and (ii) agreement of our mean relation in the hydrodynamical case with results of \citetalias{\Shao} --- give us confidence in the reliability of the NF method.

\section{Full probability distribution}\label{appx:full_pdf}

\begin{figure*}
    \centering 
    \includegraphics[width=2\columnwidth]{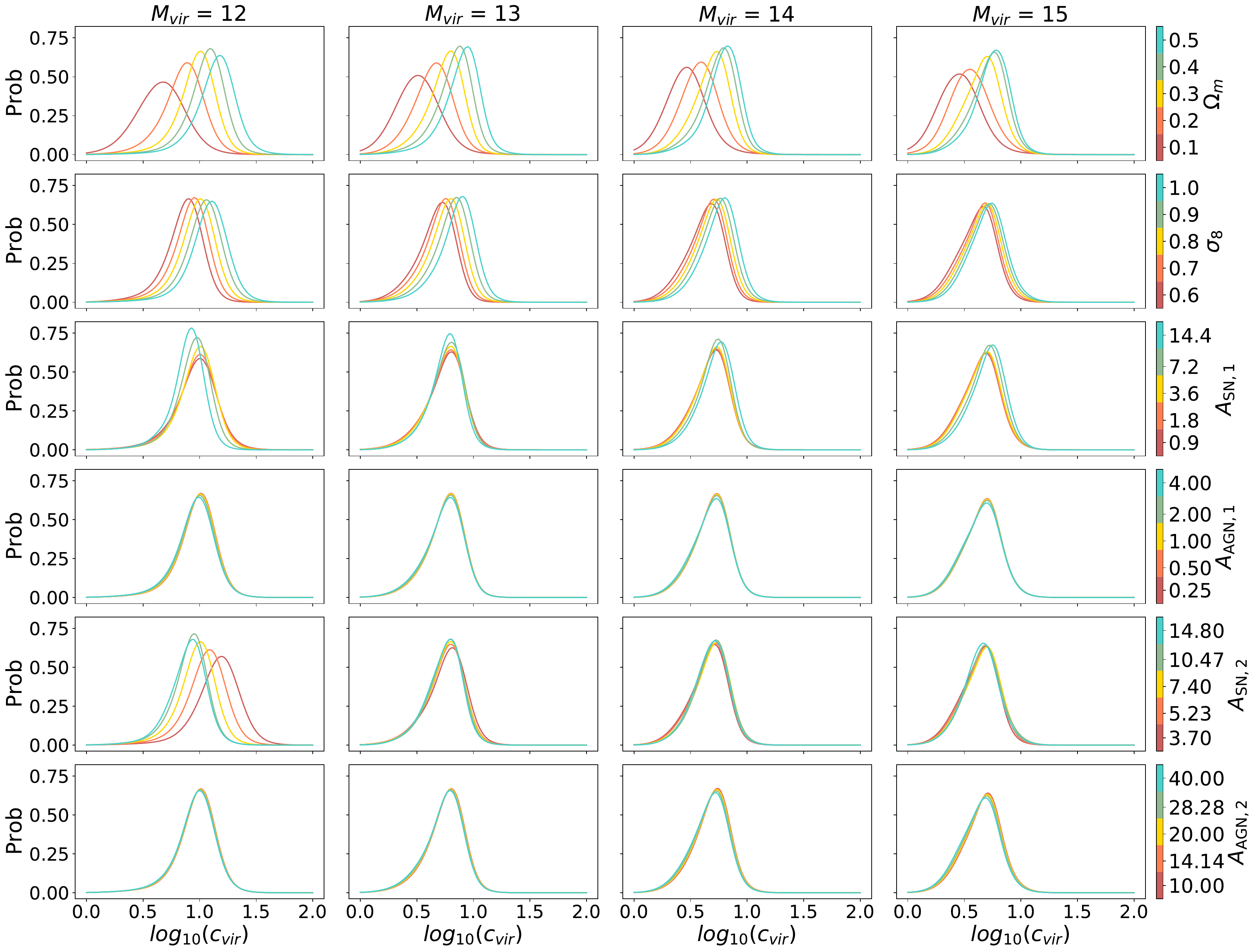}
    \caption{The full distribution of halo concentration, as predicted by the NF model. The predictions are shown for different masses (columns) and when varying different simulation parameters (rows). The results follow from the moments-based discussion of Section \ref{sec:hydro}.}
    \label{fig:PDF}
\end{figure*}

Figure \ref{fig:PDF} shows the full PDF of the concentration, when predicted by the NF model, as a function of halo mass (columns) and different parameter variations (rows). This is a more fine-grained presentation of the non-Gaussian PDF compared to the moments-based picture discussed through Figure \ref{fig:skewness} and Figure \ref{fig:kurtosis}.

\section{Results from the Astrid simulations}\label{appx:astrid}

\begin{figure*}
    \centering
    \includegraphics[width=1\linewidth]{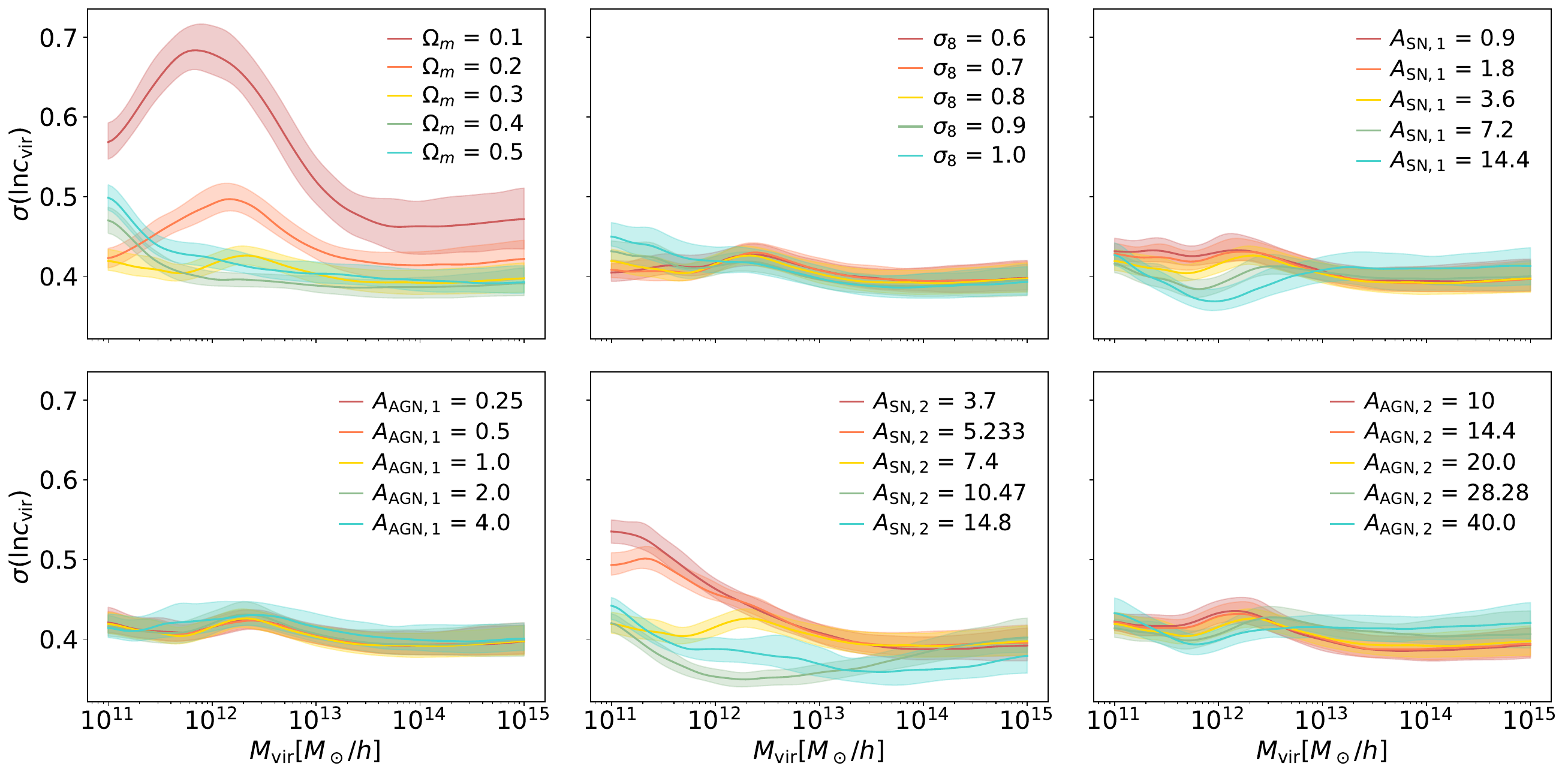}
    \caption{Similar to Figure \ref{fig:std}, but for the ASTRID model.}
    \label{fig:std_astrid}
\end{figure*}

\begin{figure*}
    \centering
    \includegraphics[width=1\linewidth]{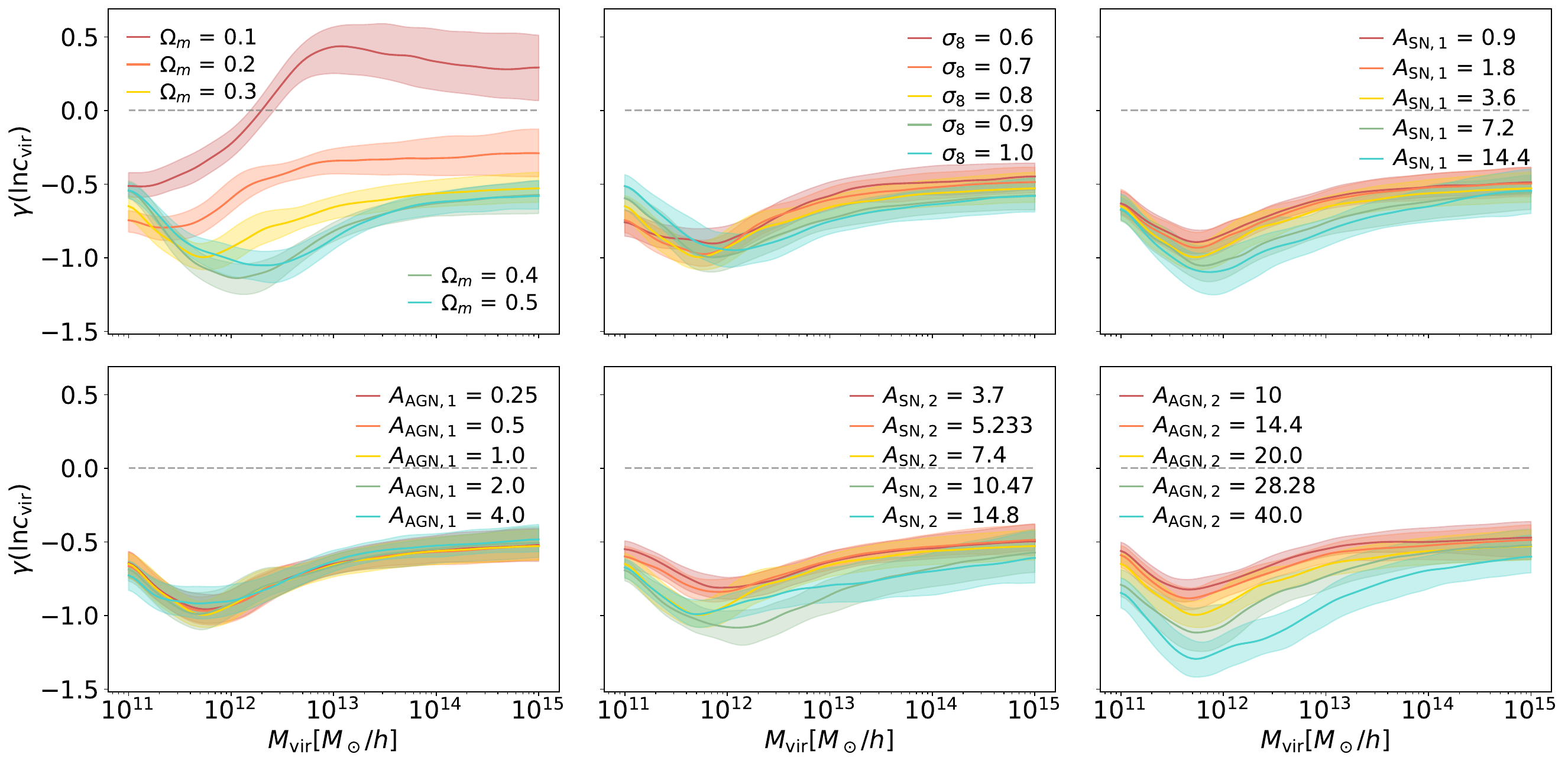}
    \caption{Similar to Figure \ref{fig:skewness}, but for the ASTRID model.}
    \label{fig:skewness_astrid}
\end{figure*}

\begin{figure*}
    \centering
    \includegraphics[width=1\linewidth]{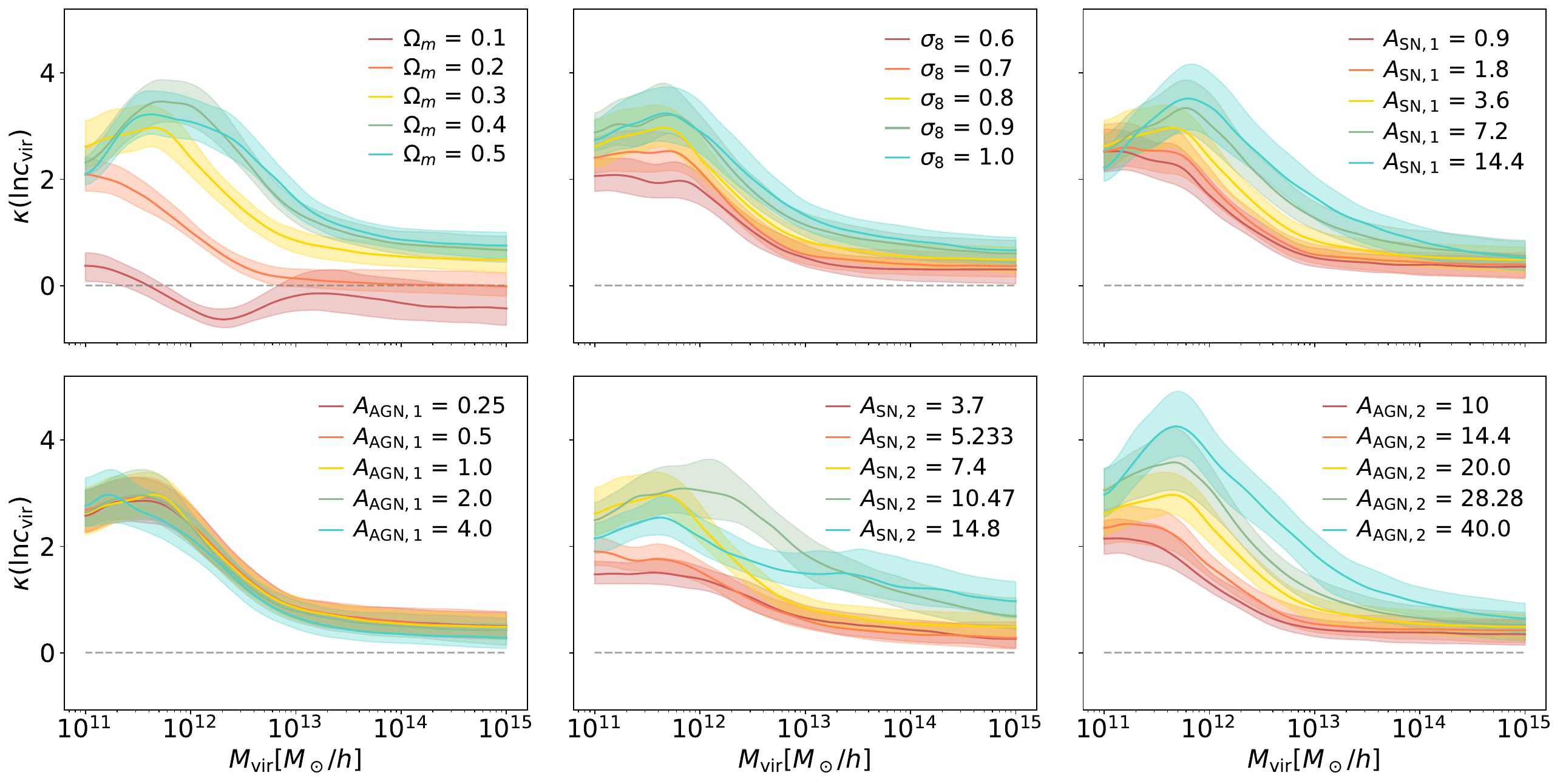}
    \caption{Similar to Figure \ref{fig:kurtosis}, but for the ASTRID model.}
    \label{fig:kurtosis_astrid}
\end{figure*}

Our main results are taken from simulations produced under the IllustrisTNG galaxy formation model. In this appendix, we extend our results to the ASTRID \citep{Bird:2022:ASTRID} galaxy formation model as well, obtained using the same procedure adopted for our main analysis with TNG. Figures \ref{fig:std_astrid}, \ref{fig:skewness_astrid}, and \ref{fig:kurtosis_astrid} present the scatter, skewness, and kurtosis. We do not discuss the model in detail as the precise meaning of the astrophysical parameters varies between TNG and ASTRID. However, we have the same structure of two SNe-focused parameters and two AGN-focused ones. For brevity, we only discuss the qualitative comparison of features across the two models.

The scatter (Figure \ref{fig:std_astrid}) has a clear feature at Milky Way-mass scales, which is amplified by increasing the baryon fraction, and can also be altered by changing the SNe feedback. This is (qualitatively) identical to the behavior in TNG. The skewness has a largely similar mass-dependent behavior to that of TNG, which reinforces the discussion in Section \ref{sec:hydro:skewness} that the skewness in the hydrodynamical case is similar to the DMO case and is therefore likely dominated by cosmic-scale processes like mergers. This skewness also exhibits a significant shift towards positive skewness as we increase the baryon fraction, once again similar to the TNG behavior in Figure \ref{fig:skewness}. Finally, the behavior of the kurtosis once again follows directly from that of the skewness, given the inequality they satisfy (see Section \ref{sec:DataMethods}).

We cannot make any generalizing statements as  we have only checked two simulations thus far (TNG and ASTRID). However, this is sufficient to say that the TNG model is not particularly unique in the features its exhibits, given these features are found in the ASTRID model as well. As discussed above, additional work on the newer simulations in \textsc{Camels} will be required to generalize these behaviors across different choices in the galaxy formation physics.

% Don't change these lines
% \bsp	% typesetting comment
\label{lastpage}
\end{document}